%% file: main.tex
\documentclass[aip,jcp,reprint,amsmath,amssymb,longbibliography,floatfix,citeautoscript]{revtex4-2}
\usepackage[T1]{fontenc} 
\usepackage[flushleft]{threeparttable}
\usepackage{array}
\usepackage{float}
\usepackage{amsmath}
\usepackage{multirow}
\usepackage{orcidlink}
\newcolumntype{P}[1]{>{\centering\arraybackslash}p{#1}}
\usepackage{pbox}

\newcommand{\wavenum}{$\text{cm}^{-1}$}
\usepackage{array}
\newcolumntype{P}[1]{>{\centering\arraybackslash}m{#1}}
\hypersetup{hidelinks}
\usepackage{pdfpages}
\makeatletter
\AtBeginDocument{\let\LS@rot\@undefined}
\makeatother  
\begin{document}
\title{
     Nuclear$-$electronic orbital second-order coupled cluster for excited states
}
\author{Jonathan H. Fetherolf\orcidlink{0000-0001-6470-1461}}
\affiliation
{Department of Chemistry, Princeton University, Princeton, NJ 08544, United States}
\author{Fabijan Pavo\v{s}evi\'c\orcidlink{0000-0002-3693-7546}}
\affiliation{Algorithmiq Ltd., FI-00160 Helsinki, Finland}
\author{Sharon Hammes-Schiffer\orcidlink{0000-0002-3782-6995}}
\email{shs566@princeton.edu}
\affiliation{Department of Chemistry, Princeton University, Princeton, NJ 08544, United States}
%
\begin{abstract}
Excited-state methods within the nuclear--electronic orbital (NEO) framework have the potential to capture vibrational, electronic, and vibronic transitions in a single calculation. In the NEO approach, specified nuclei, typically protons, are treated quantum mechanically at the same level of theory as the electrons.
Affordable excited-state NEO methods such as time-dependent density functional theory are limited to capturing the subset of excitations with single-excitation character, whereas existing methods that capture the full spectrum are limited in applicability due to their high computational cost. Herein, we introduce the excited-state variant of NEO coupled cluster with approximate second-order doubles (NEO-CC2) and its scaled-opposite-spin variant with electron-proton correlation scaling (NEO-SOS$'$-CC2). We benchmark this method for positronium hydride, where the electrons and positron are treated quantum mechanically, and find that NEO-CC2 deviates from exact results, but NEO-SOS$'$-CC2 can achieve near-quantitative accuracy by increasing the electron-positron correlation. Benchmarking NEO-CC2 and NEO-SOS$'$-CC2 on four different triatomic molecules with a quantum proton, we find that NEO-CC2 captures the correct vibrational features such as overtones and combination bands, as well as mixed electron-proton double excitations. Electron-proton correlation scaling that increases the excited-state correlation relative to the ground-state correlation improves the accuracy across all the molecular systems tested. NEO-SOS$'$-CC2 can describe single and mixed protonic and electronic excitations with accuracy approaching that of much more computationally intensive methods.
\end{abstract}
\maketitle
\section{Introduction}
Nuclear quantum effects such as tunneling, zero-point energy, proton delocalization, and quantized vibrational excited states play an essential role in many chemical and biological processes.  The nuclear--electronic orbital (NEO) framework incorporates nuclear quantum effects into electronic structure calculations by removing the Born--Oppenheimer separation between electrons and light nuclei, typically protons, treating both species at the same level of theory\cite{Webb_Iordanov_Hammes-Schiffer_2002,Pavosevic_Culpitt_Hammes-Schiffer_2020}.  Several NEO methods have been developed to accurately capture ground-state nuclear quantum effects, namely vibrational zero-point energy and proton delocalization.  NEO coupled cluster with singles and doubles (NEO-CCSD) provides excellent accuracy and systematic improvability\cite{Pavosevic_Culpitt_Hammes-Schiffer_2019,Pavosevic_Tao_Hammes-Schiffer_2021,Lambros_Fetherolf_Hammes-Schiffer_Li_2024}; however, that accuracy comes at substantial computational cost, scaling as $\mathcal{O}(N^6)$, where $N$ is a measure of system size.  Inclusion of triple excitations, particularly electron-electron-proton triples either fully (NEO-CCSDT$_\text{eep}$) or perturbatively (NEO-CCSD(T)), further improves the accuracy while increasing the computational cost, in many cases achieving quantitative agreement with experiment and reference calculations \cite{Pavosevic_Hammes-Schiffer_2022, Fowler_Brorsen_2022}.  

NEO M{\o}ller--Plesset perturbation theory is a useful alternative to costly NEO coupled-cluster methods, with a more manageable $\mathcal{O}(N^5)$ scaling \cite{Swalina_Pak_Hammes-Schiffer_2005,Fajen_Brorsen_2021}. Because electron-proton correlation can drastically alter the electronic and protonic density, orbital optimization is essential for even qualitative accuracy of the NEO wavefunction, either through variational optimization, as in orbital optimized MP2 (NEO-OOMP2),\cite{Pavosevic_Rousseau_Hammes-Schiffer_2020,Fajen_Brorsen_2020,Fetherolf_Pavosevic_Tao_Hammes-Schiffer_2022} or through the $t_1$ cluster operator, as in approximate second-order coupled cluster (NEO-CC2).\cite{Pavosevic_Hammes-Schiffer_2022} These methods are multicomponent extensions of conventional electronic OOMP2 and CC2 methods.\cite{Lochan_Head-Gordon_2007,Christiansen_Koch_Jorgensen_1995} To achieve NEO-CCSD level of accuracy, it is also necessary to compensate for a systematic underestimation of the electron-proton correlation energy due to the second-order perturbative energy. In conventional electronic structure, perturbative methods can be improved by applying a constant scaling factor in the form of spin-component scaling (SCS) or scaled-opposite-spin (SOS) of the electron correlation.\cite{Grimme_2003,Jung_Lochan_Dutoi_Head-Gordon_2004,Lochan_Head-Gordon_2007} Within the electronic SOS approximation, it is also possible to implement an $\mathcal{O}(N^4)$ routine for OOMP2 or CC2.\cite{Jung_Lochan_Dutoi_Head-Gordon_2004,Winter_Hattig_2011} The analogous NEO methods, denoted NEO-SOS$'$-OOMP2 and NEO-SOS$'$-CC2, include a scaling factor for the electron-proton correlation and have been shown to provide a similar level of accuracy as NEO-CCSD for a number of ground-state observables.\cite{Pavosevic_Rousseau_Hammes-Schiffer_2020,Fetherolf_Pavosevic_Tao_Hammes-Schiffer_2022,Pavosevic_Hammes-Schiffer_2022} Although NEO-SOS$'$-CC2 achieves similar or only slightly improved ground-state accuracy compared to NEO-SOS$'$-OOMP2, it has the added benefit of straightforward extension to excited states.

In addition to incorporating nuclear quantum effects into ground-state calculations, the NEO framework allows simultaneous computation of electronic and protonic excitations.  Electronic excitations, typically on the order of eV, are similar in accuracy to those of the underlying electronic structure method, whereas protonic excitations, on the order of~\wavenum, depend strongly on the treatment of electron-proton correlation. The most practical and widely used methods for NEO excited-state calculations are time-dependent density functional theory (NEO-TDDFT) and time-dependent Hartree-Fock (NEO-TDHF)\cite{Yang_Culpitt_Hammes-Schiffer_2018}.  With sufficiently large protonic and electronic basis sets the mean error is reduced to approximately 30~\wavenum~for the small test set studied. \cite{Culpitt_Yang_Pavosevic_Tao_Hammes-Schiffer_2019}  Although electron-proton correlation (epc) functionals\cite{Yang_Brorsen_Culpitt_Pak_Hammes-Schiffer_2017, Brorsen_Yang_Hammes-Schiffer_2017, Tao_Yang_Hammes-Schiffer_2019a} are necessary for even qualitative accuracy in ground-state NEO-DFT calculations, the quality of NEO-TDDFT protonic excitation energies does not seem to strongly depend on the use of an epc functional,\cite{Culpitt_Yang_Pavosevic_Tao_Hammes-Schiffer_2019}  presumably due to error cancellation between the ground and excited state energies. NEO-TDDFT is able to achieve good accuracy for fundamental modes when using sufficiently large basis sets, but it fails to qualitatively capture higher vibrational states such as overtones and combination bands, and it cannot capture double electron-proton vibronic excitations due to the underlying adiabatic approximation \cite{Pavosevic_Tao_Culpitt_Zhao_Li_Hammes-Schiffer_2020,Maitra_Zhang_Cave_Burke_2004}.  

Further refinement of NEO excited states requires moving from mean-field methods to a more explicit treatment of electron-electron and electron-proton correlation. High-level wavefunction methods based on equation-of-motion CCSD (NEO-EOM-CCSD) or configuration interaction (NEO-CI) improve upon NEO-TDDFT and NEO-TDHF by predicting overtones and combination bands in the correct frequency range relative to the fundamentals and capturing double electron-proton excitations.\cite{Pavosevic_Tao_Culpitt_Zhao_Li_Hammes-Schiffer_2020,Malbon_Hammes-Schiffer_2025}  The underlying electronic structure methods, EOM-CCSD or CI, carry advantages over TDDFT for computing electronic excitations in challenging single-reference problems such as charge-transfer states or large, conjugated systems \cite{Dreuw_Head-Gordon_2005}. Meanwhile, active space methods such as NEO complete active space self-consistent field (NEO-CASSCF)\cite{Webb_Iordanov_Hammes-Schiffer_2002, Fajen_Brorsen_2021a} or NEO multireference CI (NEO-MRCI)\cite{Malbon_Hammes-Schiffer_2025}  are necessary for problems with multireference character such as hydrogen tunneling.\cite{Stein_Malbon_Hammes-Schiffer_2025}  Due to their unfavorable scaling, for example $\mathcal{O}(N^6)$ for NEO-EOM-CCSD, these high-level wave function methods become prohibitively expensive for large systems.

Herein, we develop, implement, and benchmark the NEO-CC2 and NEO-SOS$'$-CC2 methods for excited states in multicomponent systems. In the context of conventional electronic structure, CC2 has gained popularity as a robust and relatively accurate low-scaling excited-state method. For electronic valence excitations, CC2 performs similarly to EOM-CCSD, with average errors of less than 0.2 eV compared to reference values.\cite{Goings_Caricato_Frisch_Li_2014, Kannar_Tajti_Szalay_2017} While CC2 struggles to capture other types of excitations, such as  Rydberg or charge transfer states, the addition of spin-component scaling corrects many of its deficiencies.\cite{Hellweg_Grun_Hattig_2008,Winter_Graf_Leutwyler_Hattig_2013,Mester_Nagy_Kallay_2017,Tajti_Szalay_2019} The scaled-opposite-spin variant is particularly attractive because it allows a further reduction in scaling from $\mathcal{O}(N^5)$ to $\mathcal{O}(N^4)$ with an improvement in accuracy compared to unscaled CC2.\cite{Winter_Hattig_2011}  Similar advantages are associated with the NEO-CC2 and NEO-SOS$'$-CC2 methods in terms of accuracy and potential for the reduction in scaling.

This paper is organized as follows. Section \ref{sec:theory} presents the theory, along with implementation details.  Section \ref{sec:results} presents the results for positronium hydride, where the electrons and positron are treated quantum mechanically, followed by the results for HeHHe$^+$, HCN, HNC, and FHF$^-$, where the proton and all electrons are treated quantum mechanically. Section \ref{sec:conclusion} presents concluding remarks.

\section{Theory}
\label{sec:theory}
\subsection{Ground-state NEO-CC2}
We begin with the normal-ordered NEO Hamiltonian in second quantization, 
\begin{equation}
    \begin{split}
        \hat{H}_\text{NEO} &= F^p_q a^q_p + \frac{1}{4}\overline{g}^{pq}_{rs}a^{rs}_{pq} \\
        &+F^P_Q a^Q_P + \frac{1}{4}\overline{g}^{PQ}_{RS}a^{RS}_{PQ} - g^{pP}_{qQ} a^{qQ}_{pP},
    \end{split}
    \label{eq:neo-ham}
\end{equation}
where $a^{q_1q_2...q_n}_{p_1p_2...p_n}=a^\dagger_{q_1}a^\dagger_{q_2}...a^\dagger_{q_n}a_{p_n}...a_{p_2}a_{p_1}$ is the general normal-ordered electronic excitation operator composed of creation and annihilation operators, $a^\dagger_p$ and $a_p$, respectively, for electronic spin orbital $p$.  $F^p_q=\langle q |\hat{F}^\text{e}|p \rangle$ are electronic Fock operator matrix elements, and $\overline{g}^{pq}_{rs}=g^{pq}_{rs}-g^{qp}_{rs}=\langle rs|pq \rangle - \langle rs|qp \rangle$ are antisymmetrized electron repulsion tensor elements.  Lowercase indices $i,j,k,l...$ denote occupied electron spin orbitals, $a,b,c,d...$ denote virtual (unoccupied) electronic spin orbitals, and $p,q,r,s...$ denote general electronic spin orbitals.  Uppercase indices are defined analogously to denote proton orbitals.  Throughout this manuscript, summation over repeated indices is assumed.

The NEO coupled cluster energy expression is 
\begin{equation}
    \begin{split}
        E_\text{NEO-CC} &= \langle0_\text{e}0_\text{p}|(1+\hat{\Lambda})e^{-\hat{T}}\hat{H}_\text{NEO} e^{\hat{T}} |0_\text{e}0_\text{p}\rangle.
    \end{split}
    \label{eq:neo-cc}
\end{equation}
Here $\hat{T} = t_\mu a^\mu$ is the excitation cluster operator and $\hat{\Lambda}=\lambda^\mu a_\mu$ is the de-excitation cluster operator, where $a^\mu = a_\mu^\dagger = \{a_i^a, a_I^A,  a_{ij}^{ab}, a_{IJ}^{AB}, a_{iI}^{aA}, ...\}$ is the set of excitation operators and $\mu$ is an excitation rank (single, double, triple, and so forth).  Specific excitation ranks such as singles and doubles will be denoted as $\mu_1$ and $\mu_2$. Moreover, $t_\mu$ and $\lambda^{\mu}$ are the unknown amplitudes that are determined by minimizing Eq. \ref{eq:neo-cc} with respect to $\lambda^{\mu}$ and $t_\mu$, respectively, yielding the $t$-amplitude and $\Lambda$-equations:
\begin{subequations}
\begin{equation}
        \frac{\partial E_\text{NEO-CC}}{\partial \lambda^\mu}=
        \langle0_\text{e}0_\text{p}|a_\mu e^{-\hat{T}}\hat{H}_\text{NEO} e^{\hat{T}}|0_\text{e}0_\text{p}\rangle = 0
\end{equation}
\label{eq:lambda}
\begin{equation}
        \frac{\partial E_\text{NEO-CC}}{\partial t_\mu}=
        \langle0_\text{e}0_\text{p}|(1+\hat{\Lambda})[e^{-\hat{T}}\hat{H}_\text{NEO} e^{\hat{T}},a^\mu]|0_\text{e}0_\text{p}\rangle = 0
\end{equation}
\label{eq:t-amp}
\end{subequations}

Previous NEO-CC work focused on singles and doubles, $\hat{T}=\hat{T}_1+\hat{T}_2$, with
\begin{subequations}
\begin{equation}
    \hat{T}_1 = t_a^i a_i^a + t_A^I a_I^A
\end{equation}
and
\begin{equation}
    \hat{T}_2 = \frac{1}{4}t_{ab}^{ij} a_{ij}^{ab} + \frac{1}{4}t_{AB}^{IJ} a_{IJ}^{AB} + t_{aA}^{iI} a_{iI}^{aA}.
\end{equation}
\end{subequations}
Inclusion of both terms as written yields NEO-CCSD\cite{Pavosevic_Culpitt_Hammes-Schiffer_2019}, inclusion of just $\hat{T}_2$ yields NEO-CCD\cite{Pavosevic_Rousseau_Hammes-Schiffer_2020}, and augmenting NEO-CCSD with the electron-electron-proton triples term yields $\hat{T}_3^\text{eep}=t_{abA}^{ijI} a_{ijI}^{abA}$ gives NEO-CCSDT$_\text{eep}$.\cite{Pavosevic_Hammes-Schiffer_2022}
Utilizing $\hat{T}_1$ similarity-transformed operators $\bar{\mathcal{O}}=e^{-\hat{T}_1}\mathcal{O}e^{\hat{T}_1}$, the expressions for the $t_1$ and $t_2$ amplitudes at the NEO-CCSD level are
\begin{subequations}
\begin{equation}
\Omega_{\mu_1}=\langle0_\text{e}0_\text{p}|a_{\mu_1}(\bar{H}+[\bar{H},\hat{T}_2])|0_\text{e}0_\text{p}\rangle = 0
\label{eq:ccsdsingles}
\end{equation}
and
\begin{equation}
\Omega_{\mu_2} = \langle0_\text{e}0_\text{p}|a_{\mu_2}(\bar{H}+[\bar{H},\hat{T}_2]
+\frac{1}{2}[[\bar{H},\hat{T}_2],\hat{T}_2])|0_\text{e}0_\text{p}\rangle = 0
\label{eq:ccsddoubles}
\end{equation}
\end{subequations}

NEO-CC2 retains the full $t_1$ amplitude expression from NEO-CCSD, Eq. \ref{eq:ccsdsingles}.  However, the $t_2$ amplitude expression is truncated at second order, giving
\begin{equation}
\Omega_{\mu_2}=\langle0_\text{e}0_\text{p}|a_{\mu_2}(\bar{H}+[\hat{F},\hat{T}_2])|0_\text{e}0_\text{p}\rangle = 0
\label{eq:ccsddoubles}
\end{equation}
where $\hat{F}$ is the NEO Fock operator, $F^p_q a^q_p + F^P_Q a^Q_P$.

\subsection{Extension to excited states}
Excited states of the NEO-CC2 Hamiltonian are obtained within coupled cluster response theory\cite{Koch_Jo/rgensen_1990,Pedersen_Koch_1997} via diagonalization of the Jacobian matrix
\begin{equation}
\begin{aligned}
\mathbf{A}_{\mu_i\nu_j} &= \frac{\partial\Omega_{\mu_i}}{\partial a_{\nu_j}} = \begin{pmatrix}
\mathbf{A}_{\mu_1\nu_1} & \mathbf{A}_{\mu_1\nu_2} \\
\mathbf{A}_{\mu_2\nu_1} & \mathbf{A}_{\mu_2\nu_2}
\end{pmatrix}
\end{aligned}
\end{equation}
composed of the singles-singles, singles-doubles, doubles-singles and doubles-doubles blocks. These blocks can be respectively decomposed into terms involving $t_1^\text{e}$, $t_1^\text{p}$, $t_2^\text{ee}$, and $t_2^\text{ep}$:
\begin{subequations}
    \begin{equation}
        \mathbf{A}_{\mu_1\nu_1} = \begin{pmatrix}
\mathbf{A}_{\mu_1^\text{e}\nu_1^\text{e}} & \mathbf{A}_{\mu_1^\text{e}\nu_1^\text{p}} \\
\mathbf{A}_{\mu_1^\text{p}\nu_1^\text{e}} & \mathbf{A}_{\mu_1^\text{p}\nu_1^\text{p}}
\end{pmatrix},
    \end{equation}
        \begin{equation}
        \mathbf{A}_{\mu_1\nu_2} = \begin{pmatrix}
\mathbf{A}_{\mu_1^\text{e}\nu_2^\text{ee}} & \mathbf{A}_{\mu_1^\text{e}\nu_2^\text{ep}} \\
\mathbf{A}_{\mu_1^\text{p}\nu_2^\text{ee}} & \mathbf{A}_{\mu_1^\text{p}\nu_2^\text{ep}}
\end{pmatrix},
    \end{equation}
        \begin{equation}
        \mathbf{A}_{\mu_2\nu_1} = \begin{pmatrix}
\mathbf{A}_{\mu_2^\text{ee}\nu_1^\text{e}} & \mathbf{A}_{\mu_2^\text{ee}\nu_1^\text{p}} \\
\mathbf{A}_{\mu_2^\text{ep}\nu_1^\text{e}} & \mathbf{A}_{\mu_2^\text{ep}\nu_1^\text{p}}
\end{pmatrix},
    \end{equation}
        \begin{equation}
        \mathbf{A}_{\mu_2\nu_2} = \begin{pmatrix}
\mathbf{A}_{\mu_2^\text{ee}\nu_2^\text{ee}} & \mathbf{A}_{\mu_2^\text{ee}\nu_2^\text{ep}} \\
\mathbf{A}_{\mu_2^\text{ep}\nu_2^\text{ee}} & \mathbf{A}_{\mu_2^\text{ep}\nu_2^\text{ep}}
\end{pmatrix}.
    \end{equation}
\label{eq:matrix_elements}
\end{subequations}
This work focuses on systems with one quantum proton and thus does not require  proton-proton correlation (i.e., no $t_2^\text{pp}$).  Inclusion of such terms would involve a straightforward expansion of $\mathbf{A}_{\mu_1\nu_2}$, $\mathbf{A}_{\mu_2\nu_1}$, and $\mathbf{A}_{\mu_2\nu_2}$ from $2\times2$ block matrices to $2\times3$, $3\times2$ and $3\times3$ block matrices, respectively, to accommodate blocks involving $\mu_2^\text{pp}$ and $\nu_2^\text{pp}$. Programmable equations for all sixteen matrix elements in Eq. \ref{eq:matrix_elements} are given in the Supplementary Material.

To avoid the costly step of constructing and diagonalizing the entire $\mathbf{A}_{\mu_i\nu_j}$ matrix, we use the L\"owdin partitioning technique\cite{Lowdin_1982}. This approach reduces the Jacobian from the full single- and double-excitation space to the much smaller single-excitation space, represented by an effective Jacobian,
\begin{equation}
    \mathbf{A}^\text{eff}_{\mu_1\nu_1}(\omega) =
\mathbf{A}_{\mu_1\nu_1}-\frac{\mathbf{A}_{\mu_1\gamma_2}\mathbf{A}_{\gamma_2\nu_1}}{\varepsilon_{\gamma_2}-\omega},
\end{equation}
Here $\varepsilon_{\gamma_2}=\varepsilon_a - \varepsilon_i + \varepsilon_b - \varepsilon_j$ is the orbital energy difference between occupied orbitals $i,j$ and virtual orbitals $a,b$ composing the double excitation $\gamma_2$, and $\omega$ is the unknown excitation energy that is obtained by inserting $\mathbf{A}^\text{eff}_{\mu_1\nu_1}(\omega)$ into the nonlinear effective eigenvalue problem, 
\begin{equation}
\mathbf{A}^\text{eff}_{\mu_1\nu_1}(\omega)E_{\nu_1} = \omega E_{\mu_1}.
\label{eq:lowdin_eig}
\end{equation}
With convergence of Eq. \ref{eq:lowdin_eig},  $E_{\nu_1}=E_{\mu_1}$ will be the single-excitation portion of the eigenvector of the $\mathbf{A}_{\mu_i\nu_j}$ corresponding to the eigenvalue of $\omega$. Eigenvectors associated with double excitations can be obtained from the single-excitation manifold with
\begin{equation}
    E_{\mu_2}=-\frac{\mathbf{A}_{\mu_2\nu_1}E_{\nu_1}}{\varepsilon_{\mu_2}-\omega}.
\end{equation}

Following previous work on scaled-opposite-spin (SOS) CC2\cite{Winter_Hattig_2011,Tajti_Szalay_2019,Hellweg_Grun_Hattig_2008}, we separately scale the ground and excited state correlation contributions of $\mathbf{A}^\text{eff}_{\mu_1\nu_1}$:
\begin{equation}
\mathbf{A}_{\mu_1\nu_1} = \begin{pmatrix}
c_\text{os}^\text{gs}\mathbf{A}_{\mu_1^\text{e}\nu_1^\text{e}} & c_\text{ep}^\text{gs}\mathbf{A}_{\mu_1^\text{e}\nu_1^\text{p}} \\
c_\text{ep}^\text{gs}\mathbf{A}_{\mu_1^\text{p}\nu_1^\text{e}} & \mathbf{A}_{\mu_1^\text{p}\nu_1^\text{p}}
\end{pmatrix}
\label{eq:gsscaling}
\end{equation}
and
\begin{equation}
\frac{\mathbf{A}_{\mu_1\gamma_2}\mathbf{A}_{\gamma_2\nu_1}}{\varepsilon_{\gamma_2}-\omega}=\mathbf{A}^{'}_{\mu_1\nu_1}=
\begin{pmatrix}
c_\text{os}^\text{ex}\mathbf{A}^{'}_{\mu_1^\text{e}\nu_1^\text{e}} & c_\text{ep}^\text{ex}\mathbf{A}^{'}_{\mu_1^\text{e}\nu_1^\text{p}} \\
c_\text{ep}^\text{ex}\mathbf{A}^{'}_{\mu_1^\text{p}\nu_1^\text{e}} & \mathbf{A}^{'}_{\mu_1^\text{p}\nu_1^\text{p}}
\label{eq:exscaling}
\end{pmatrix}.
\end{equation}
The same-spin electronic components of $\mathbf{A}_{\mu_1^\text{e}\nu_1^\text{e}}$ and $\mathbf{A}^{'}_{\mu_1^\text{e}\nu_1^\text{e}}$ are set to zero in keeping with the scaled-opposite-spin approach.

\subsection{Implementation details}
The excited state NEO-CC2 and NEO-SOS$'$-CC2 methods are implemented in Q-Chem 6.3.\cite{Epifanovsky_Gilbert_Feng_Lee_Mao_Mardirossian_etal_2021} For the purposes of benchmarking, the method was implemented without density fitting and for only a single quantum proton. When computing the initial NEO-HF orbitals with a single quantum proton, an alternative molecular Hamiltonian that omits the proton-proton interaction $V_\text{pp}$ is used.
Although the $V_\text{pp}$ term has no effect on the NEO-HF energy for a single proton, keeping it would produce a different set of unoccupied orbitals.\cite{Swalina_Pak_Hammes-Schiffer_2005}

\section{Results and Discussion}
\label{sec:results}
\subsection{Positronium hydride}
As a first test, we compute the excitation energies of positronium hydride (PsH).  PsH consists of a positron (Ps$^+$) bound to a hydride (H$^-$) ion made up of a proton and two electrons. In these calculations, we treat the hydrogen nucleus classically (as a point charge), while the electrons and positron are treated quantum mechanically. PsH has been used in the past as a benchmark for several multicomponent excited-state quantum chemistry methods.\cite{Tachikawa_2001, Adamson_Duan_Burggraf_Pak_Swalina_Hammes-Schiffer_2008, Pavosevic_Hammes-Schiffer_2019}
NEO-CC2 and NEO-SOS$'$-CC2 calculations were performed using identical augmented correlation-consistent Dunning basis sets for both the electrons and positron.  For both basis sets we compare to NEO-EOM-CCSD, and for the smaller aug-cc-pVTZ basis set, we also benchmark against the numerically exact NEO full configuration-interaction (NEO-FCI).\cite{Pavosevic_Hammes-Schiffer_2019} For NEO-SOS$'$-CC2, we use an electron-positron scaling factor $c^
\text{gs}_\text{ep}=c^\text{ex}_\text{ep}=1.3$, the same value that is typically used for the opposite-spin electron-electron scaling factors $c^\text{gs}_\text{os}$ and $c^\text{ex}_\text{os}$.\cite{Lochan_Head-Gordon_2007,Winter_Hattig_2011}
\begin{table*}[ht]
\caption{Ground state correlation energy and first three excitation energies of PsH.  Results are calculated with the specified methods and either the aug-cc-pVTZ (aTZ) or aug-cc-pV5Z (a5Z) basis set for both the electrons and positron.  All values are given in hartree.}
\begin{threeparttable}
\input{tables/psh_table}
\begin{tablenotes}
\item[a] Obtained from Ref. \citenum{Pavosevic_Hammes-Schiffer_2019}.
\item[b] Computed using SOS$'$ scaling factors $(c^\text{gs}_\text{ep}=1.3,c^\text{ex}_\text{ep}=1.3)$
\item[c] Mean unsigned error relative to NEO-FCI.
\item[d] Mean unsigned error relative to NEO-EOM-CCSD.
\end{tablenotes}
\end{threeparttable}
\label{tab:psh}
\end{table*}

We present results for the ground state correlation energy ($E_\text{corr}$) and the first three excited states of PsH in Table \ref{tab:psh}.  $E_\text{corr}$ includes both electron-electron and electron-positron correlation energy.  We first focus on the aug-cc-pVTZ results. Compared to the NEO-FCI benchmark, NEO-EOM-CCSD achieves high accuracy, with a mean unsigned error (MUE) of 2.2 mHa. Unscaled NEO-CC2 results in a relative large MUE of 11.3 mHa compared to NEO-FCI. NEO-CC2 struggles particularly with capturing the $E_\text{corr}$ of PsH, which is underestimated by 25 mHa. NEO-SOS$'$-CC2, on the other hand, performs almost as well as NEO-EOM-CCSD, with a MUE of 3.0 mHa relative to NEO-FCI. Compared to NEO-CC2, the SOS$'$ drastically improves $E_\text{corr}$, which is now similar to NEO-CCSD, while also leading to overall improved excitation energies.

We computed the $E_\text{corr}$ and excitation energies of PsH with the aug-cc-pV5Z basis set to test if NEO-CC2 and NEO-SOS$'$-CC2 exhibits similar basis set convergence to NEO-EOM-CCSD. In this case, we use NEO-EOM-CCSD as the benchmark. We again see that NEO-CC2 produces a rather large MUE, in this case 14.1 mHa. Again, NEO-CC2 struggles to capture the full correlation energy $E_\text{corr}$ but also produces fairly large errors in the excitation energies. NEO-SOS$'$-CC2 differs from NEO-EOM-CCSD by a MUE of only 2.8 mHa.

In this section, we have shown that NEO-SOS$'$-CC2 can accurately predict both the ground and excited state energies of PsH. The accuracy of NEO-SOS$'$-CC2 is similar to that of NEO-EOM-CCSD when compared to NEO-FCI for a smaller basis set and exhibits similar convergence with basis set size. NEO-CC2 performed markedly worse than all other methods considered and was particularly hampered by an underestimation of the electron-positron correlation energy.
\subsection{HeHHe$^+$}
In the previous section, we demonstrated that NEO-SOS$'$-CC2 can predict the ground and excited state energies of PsH at a level of accuracy that is competitive with much more computationally demanding methods.  For the remainder of this paper, we will focus on systems involving electrons and quantum protons. Such systems present additional theoretical challenges due to the disparate masses and corresponding localization properties of the electron and proton wave functions. With this in mind, we adopt the $\gamma$-def2-QZVP* electronic basis sets from Ref. \citenum{Malbon_Hammes-Schiffer_2025}, which were developed to combat over-stabilization of the protonic ground state relative to excited states in NEO-CI methods. In these calculations, the heavy nuclei are fixed, and therefore the computed excitation energies are not directly comparable to experimental measurements. 
\begin{table*}[ht]
\caption{Ground state energy and first excitation energy for HeHHe$^+$. The NEO-SOS$'$-CC2 results are given with different scaling factors as defined in Eqs. \ref{eq:gsscaling} and \ref{eq:exscaling}, reported as $(c^\text{gs}_\text{ep},c^\text{ex}_\text{ep})$.}
\begin{threeparttable}
\input{tables/hehhe_fci}
\begin{tablenotes}
\item[a] Obtained from Ref. \citenum{Malbon_Hammes-Schiffer_2025}.
\item[b] Computed with 6-31G/0.4404-def2-QZVP*/8s8p8d8f basis set.
\item[c] Computed with 6-31G/cc-pV6Z/8s8p8d8f basis set. $E_0$ is not shown due to basis set discrepancy.
\end{tablenotes}
\end{threeparttable}
\label{tab:hehhe_fci}
\end{table*}

We first test NEO-CC2 on the HeHHe$^+$ molecule, where the two He nuclei are fixed and the central hydrogen nucleus is treated quantum mechanically. This molecule has been used as a benchmark system for several multicomponent excited-state methods.\cite{Skone_Pak_Hammes-Schiffer_2005,Alaal_Brorsen_2021,Feldmann_Muolo_Baiardi_Reiher_2022,Garner_Upadhyay_Li_Hammes-Schiffer_2024,Garner_Upadhyay_Li_Hammes-Schiffer_2025a,Malbon_Hammes-Schiffer_2025} Because it contains only four electrons and one quantum nucleus, it is possible to compute the NEO-FCI solution for HeHHe$^+$ with a modestly sized basis set. In Table \ref{tab:hehhe_fci} we report the total ground-state energy $E_0$ and first excitation energy $E_1-E_0$ for HeHHe$^+$ using the 6-31G electronic basis set\cite{Ditchfield_Hehre_Pople_1971} on the helium atoms, the  0.4404-def2-QZVP* electronic basis set\cite{Malbon_Hammes-Schiffer_2025} on the hydrogen, and an even-tempered 8s8p8d8f protonic basis set with exponents from $2\sqrt{2}$ to 32.\cite{Bardo_Ruedenberg_1974, Yang_Brorsen_Culpitt_Pak_Hammes-Schiffer_2017} For all HeHHe$^+$ calculations, the He-He distance is set at 1.8 \AA. These parameters allow us to compare directly to NEO-FCI results published in Ref. \citenum{Malbon_Hammes-Schiffer_2025}. We also compare to the Fourier Grid Hamiltonian (FGH) method, which provides numerically exact vibrational energy levels for the electronic potential energy surface within the adiabatic approximation.\cite{Marston_Balint-Kurti_1989,Webb_Hammes-Schiffer_2000a} The FGH results used herein were obtained with the electronic grid calculated at the conventional CCSD level of theory with the aug-cc-pVTZ electronic basis set. We also compare to NEO-TDDFT calculations using the B3LYP electronic exchange-correlation functional\cite{Lee_Yang_Parr_1988,Becke_1993} and the epc17-2 electron-proton correlation functional.\cite{Yang_Brorsen_Culpitt_Pak_Hammes-Schiffer_2017, Brorsen_Yang_Hammes-Schiffer_2017} Note that Cartesian rather than spherical Gaussian functions were used in both the electronic and protonic basis sets for NEO-TDDFT to maintain consistency with previous work. \cite{Culpitt_Yang_Pavosevic_Tao_Hammes-Schiffer_2019}  All other calculations in this work used spherical Gaussian functions.

NEO-CC2 produces a ground state energy that is around 15 mHa higher than the energy obtained with NEO-FCI. Clearly an approximate method such as CC2 will not converge to the total FCI correlation energy, and a comparison of absolute ground state energies between methods is not meaningful. Nonetheless, we include ground-state energies to track the effect of SOS$'$ scaling factors on the NEO-CC2 energy. More importantly, NEO-CC2 overestimates the first protonic excited state $E_1$, corresponding to the hydrogen bending mode, by around 200 \wavenum~compared to FGH and NEO-FCI. This error is comparable to that of NEO-TDDFT with a large cc-pV6Z basis set on the hydrogen. However, while NEO-TDDFT underestimates the bending frequency, NEO-CC2 overestimates the excitation energy. The error in NEO-CC2 can be attributed to the tendency of post-HF NEO methods to capture electron-proton correlation more readily for the ground state than for excited states, leading to overestimation of the protonic excitation energies. The NEO-TDDFT formalism, on the other hand, leads to a more effective cancellation of error between the ground and excited state energies.

The uneven stabilization of ground- versus excited-state correlation is even more evident when we explore the effect of SOS$'$ electron-proton correlation scaling.  Scaling factors are reported as ($c^\text{gs}_\text{ep}$,$c^\text{ex}_\text{ep}$) throughout this discussion. Although in PsH the standard SOS scaling factor of (1.3,1.3) notably improved both the ground- and excited-state accuracy, this is not the case for HeHHe$^{+}$.  Table \ref{tab:hehhe_fci} shows that with these scaling factors, the ground-state energy is lowered to below the NEO-FCI value, but the excitation energy increases by more than 1000 \wavenum. Any value of $c^\text{gs}_\text{ep}>1$ leads to a disproportionate stabilization of the ground state and degrades the accuracy of the energy gap. This issue can be addressed by separately reducing $c^\text{gs}_\text{ep}$ or increasing $c^\text{ex}_\text{ep}$ relative to one another. We find the values of (0.91,1.0) or (1.0,8.0) lead to optimal agreement with the NEO-FCI and FGH excitation energies, with relatively minimal effect on the ground state energy (Table \ref{tab:hehhe_fci}). Because the two approaches lead to similar performance, we will use (0.91,1.0) scaling factors in the main paper and report the corresponding (1.0,8.0) results in the Supplementary Material.
%
\begin{table*}[ht]
\caption{Energies of lowest-lying excited states for HeHHe$^+$. The states are labeled according to quanta in the protonic vibrational modes ($\nu_1$,$\nu_2$,$\nu_3$), where $\nu_1$ and $\nu_2$ are degenerate bending modes and $\nu_3$ is the stretching mode. The NEO-SOS$'$-CC2 results are given with scaling factors $c^\text{gs}_\text{ep}=0.91$ and $c^\text{ex}_\text{ep}=1.0$. All energies are given in \wavenum.}
\begin{threeparttable}
\input{tables/hehhe}
\begin{tablenotes}
\item[a] Obtained from Ref. \citenum{Malbon_Hammes-Schiffer_2025}.
\item[b] Computed with aug-cc-pVTZ/0.4404-def2-QZVP*/8s8p8d8f8g basis set.
\item[c] Computed with aug-cc-pVTZ/cc-pV6Z/8s8p8d8f8g basis set.
\end{tablenotes}
\end{threeparttable}
\label{tab:hehhe_tz}
\end{table*}

We will now explore how NEO-SOS$'$-CC2 captures the proton vibrational spectrum, namely the higher protonic vibrational excitations, of HeHHe$^+$. For these calculations, we use the aug-cc-pVTZ electronic basis set on the heavy nuclei, the 0.4404-def2-QZVP* electronic basis set on the quantum hydrogen, and the 8s8p8d8f8g protonic basis set. Table \ref{tab:hehhe_tz} provides the frequencies for the first four protonic excitations, denoted according to the quanta in the fundamental protonic vibrations ($\nu_1$,$\nu_2$,$\nu_3$), where $\nu_1$ and $\nu_2$ are degenerate bending modes and $\nu_3$ is the stretching mode. We compare to numerically exact FGH reference calculations, as well as high-level NEO-MRCI calculations with singles, doubles, and electron-electron-proton triples (NEO-MR-SDT$_\text{een}$CI) from Ref. \citenum{Malbon_Hammes-Schiffer_2025}. The FGH method predicts the degenerate bending mode of HeHHe$^+$ to be 772 \wavenum~and the stretching mode to be 1818 \wavenum. NEO-MR-SDT$_\text{een}$CI gives qualitatively similar results with frequencies of 723 and 2371 \wavenum, respectively. Crucially, both the FGH and NEO-MR-SDT$_\text{een}$CI methods produce overtones and combination bands of the degenerate bend at roughly twice the bending frequency, with values between the bending and stretching mode frequencies. Although NEO-TDDFT can predict relatively accurate fundamental vibrational frequencies, the overtones and combination bands are always far too high in energy.\cite{Culpitt_Yang_Pavosevic_Tao_Hammes-Schiffer_2019,Pavosevic_Tao_Culpitt_Zhao_Li_Hammes-Schiffer_2020} This behavior is demonstrated in Table \ref{tab:hehhe_tz}, which shows that NEO-TDDFT predicts overtones and combination bands that are nearly ten times the bending frequency.

As with the smaller basis set, unscaled NEO-CC2 overestimates the bending mode frequency by around 200 \wavenum, whereas NEO-MR-SDT$_\text{een}$CI is within 50 \wavenum~of the FGH reference. The NEO-CC2 stretching frequency is overestimated as well but is of similar accuracy as that predicted by NEO-MR-SDT$_\text{een}$CI. The FGH method predicts the overtones and combination band to be around 2.05 and 2.1 times the fundamental bending frequency, respectively, whereas NEO-MR-SDT$_\text{een}$CI predicts these frequencies to be around 2.3 and 2.5 times the fundamental bending frequency. NEO-CC2 predicts these frequencies to be around 2.3 times and 2.4 times the bending frequency, in very close agreement with NEO-MR-SDT$_\text{een}$CI. Because the bending frequency is overestimated by NEO-CC2, the overtones and combination bands are not quite below the stretch, resulting in a qualitative error in the state ordering compared to the FGH and NEO-MR-SDT$_\text{een}$CI ordering. 

NEO-SOS$'$-CC2 predicts vibrational frequencies with higher accuracy than NEO-TDDFT and only slightly lower accuracy than the much more costly NEO-MR-SDT$_\text{een}$CI. In this case, we use the SOS$'$ electron-proton correlation scaling factors that were tuned to give the NEO-FCI bending frequency. As expected, these scaling factors remain valid for the larger basis set, leading to nearly quantitative agreement with the FGH bending mode frequency. Additionally, the stretching mode is lowered in energy, bringing it closer to the FGH result. The overtones and combination band are also lowered in energy, becoming close to or slightly lower than the stretching mode. Thus, we have shown that NEO-CC2 captures several important features of the proton vibrational spectrum of HeHHe$^+$, including overtones and combination bands, while electron-proton correlation scaling  further improves the accuracy of all modes when compared to the FGH reference.
\subsection{HCN}
As a further test of NEO-SOS$'$-CC2, we compute the protonic and vibronic excitation energies of hydrogen cyanide, HCN. In contrast to the internal hydrogen of HeHHe$^+$, the terminal hydrogen of HCN feels a softer, asymmetric potential along the bonding axis. As a result, the proton delocalization and the low-frequency bending mode have historically been more difficult to capture with NEO methods.\cite{Pavosevic_Tao_Culpitt_Zhao_Li_Hammes-Schiffer_2020,Alaal_Brorsen_2021,Pavosevic_Hammes-Schiffer_2022} In Table \ref{tab:hcn_tz}, we examine the first four excited states of HCN. We compare our results to FGH reference frequencies from Ref. \citenum{Pavosevic_Tao_Culpitt_Zhao_Li_Hammes-Schiffer_2020}, where the electronic grid was calculated at the CCSD/aug-cc-pVTZ level of theory. The FGH method predicts a degenerate bending mode at 634 \wavenum~and a stretching mode at 3173 \wavenum. Similar to HeHHe$^+$, HCN exhibits an overtone and combination band that are both lower in frequency than the stretching mode, occurring at 1277 and 1261 \wavenum, respectively. 
\begin{table*}[!htbp]
\caption{First four protonic excited states of HCN. The NEO-SOS$'$-CC2 results are given with scaling factors $c^\text{gs}_\text{ep}=0.91$ and $c^\text{ex}_\text{ep}=1.0$. All values are given in \wavenum.}
\begin{threeparttable}
\input{tables/hcn}
\begin{tablenotes}
\item[a] Obtained from Ref. \citenum{Pavosevic_Tao_Culpitt_Zhao_Li_Hammes-Schiffer_2020}.
\item[b] Obtained from Ref. \citenum{Malbon_Hammes-Schiffer_2025}.
\item[c] Computed using cc-pVTZ/0.1393-def2-QZVP*/8s8p8d8f basis.
\item[d] Computed using cc-pVTZ/cc-pV6Z/8s8p8d8f basis.
\end{tablenotes}
\end{threeparttable}
\label{tab:hcn_tz}
\end{table*}

We also compare our results to NEO-MR-SDT$_\text{een}$CI results from Ref. \citenum{Malbon_Hammes-Schiffer_2025}, which uses the $\gamma$-def2-QZVP* electronic basis set with $\gamma=0.1393$ for the terminal hydrogen. The NEO-MR-SDT$_\text{een}$CI fundamental bending and stretching modes occur at 678 and 2968 \wavenum, respectively, in good agreement with the FGH reference. Although NEO-MR-SDT$_\text{een}$CI values are not reported for overtones and combination band frequencies, the paper indicates that four excitations occur between the bending and stretching modes, corresponding to all double and triple occupations of the bending mode. We also include NEO-TDDFT results, which were computed using the cc-pV6Z electronic basis set on the quantum proton. In these results, NEO-TDDFT shows very good agreement with the FGH benchmark for the fundamental modes, but the overtones and combination band are much too high in energy, occurring at around ten times the fundamental bending frequency.

Our NEO-CC2 and NEO-SOS$'$-CC2 calculations were performed with the same cc-pVTZ/0.1393-def2-QZVP*/8s8p8d8f basis sets and HCN geometry as those used for the NEO-MR-SDT$_\text{een}$CI calculations reported in Ref. \citenum{Malbon_Hammes-Schiffer_2025}. With no scaling factors, NEO-CC2 predicts fundamental bending and stretching frequencies of 1223 and 3121 \wavenum, while the overtones and combination band occur at 2970 and 3038 \wavenum, respectively. The stretching mode is in good agreement with the FGH reference, but the bending mode is 500 \wavenum~too high in frequency. The overtones and combination band are much more reasonable than those obtained with NEO-TDDFT, occurring at 2.4 and 2.5 times the bending frequency, respectively. Again, we performed NEO-SOS$'$-CC2 calculations using scaling factors $c^\text{gs}_\text{ep}=0.91$ and $c^\text{ex}_\text{ep}=1.0$. With SOS$'$ scaling, the bending mode frequency is brought into more reasonable agreement with the FGH reference, while the stretching mode frequency is below the reference value by around 400 \wavenum. Crucially, both NEO-CC2 and NEO-SOS$'$-CC2 predict the correct energetic ordering of the four modes examined. While electron-proton correlation scaling can be used to bring the bending mode frequency into better agreement with reference values, HCN remains a challenging system for NEO methods to achieve simultaneous quantitative accuracy for the fundamental bending mode and stretching mode frequencies. In the Supplementary Material Table S1, we include a comparison of the excited states of HCN computed by different NEO methods using the Dunning correlation consistent basis sets rather than the NEO-specific $\gamma$-def2-QZVP* basis set, allowing a direct comparison to NEO-EOM-CCSD results from Ref. \citenum{Pavosevic_Tao_Culpitt_Zhao_Li_Hammes-Schiffer_2020}. We observe that NEO-CC2 is qualitatively similar but less accurate than NEO-EOM-CCSD for all basis sets, whereas NEO-SOS$'$-CC2, with no additional tuning of the electron-proton scaling factors, is close in accuracy to NEO-EOM-CCSD.

Finally, we examine the double-excitation manifold for the NEO-SOS$'$-CC2 excitation spectrum. Although this spectrum could include doubly excited electronic states, or doubly excited protonic states in the case of multiple protons, we will focus here on mixed electron-proton vibronic double excitations. In Table \ref{tab:hcn_vibron}, we present the lowest-lying series of vibronic excitations for HCN computed using the cc-pVTZ/0.1393-def2-QZVP*/8s8p8d8f basis sets and SOS$'$ scaling factors $c^\text{gs}_\text{ep}=0.91$ and $c^\text{ex}_\text{ep}=1.0$. The corresponding electronic excitation occurs at 9.82 eV, while the first mixed electron-proton excitation occurs at 14.03 eV and is assigned as the simultaneous excitation of the electron and the protonic bending mode, $e_1p_{010}/e_1p_{100}$. Further double excitations occur at 14.25, 14.38 and 14.39 eV, corresponding to the simultaneous excitation of the electron and the stretching mode, the electron and the combination band, and the electron and the bending overtones, respectively. The large gap between the bare electronic excitation and the mixed electron-proton excitations can be attributed to the truncation of the coupled cluster $T$ operator, such that it will provide a much better correlation description in the single excitation manifold. A similar trend is observed in results for the electron-proton excitation of HCN using NEO-EOM-CCSD, as obtained from Ref. \citenum{Pavosevic_Tao_Culpitt_Zhao_Li_Hammes-Schiffer_2020}. In this case, the bare electronic excitation occurs at 9.71 eV, while the first mixed electron-proton excitation is 4.21 eV higher, at 13.92 eV. The same gap of 4.21 eV occurs between $e_1p_{000}$ at 9.82 eV and $e_1p_{100}/e_1p_{010}$ at 14.03 with NEO-SOS$'$-CC2. Similarly, the gaps between the first two mixed electron-proton excitations are 0.20 eV for NEO-EOM-CCSD versus 0.22 eV for NEO-SOS$'$-CC2, indicating that the vibronic spectra have a similar structure with a constant shift of around 0.1 eV.
\begin{table*}[ht]
\caption{Lowest-lying electron-proton double excitations of HCN. The electronic excited state is denoted as $e_1$, and the protonic state is denoted as $p_{n_1n_2n_3}$ , where the subscript indices give the number of quanta in the degenerate bending modes $\nu_1$ and $\nu_2$ and stretching mode $\nu_3$. All values are given in eV.}
\begin{threeparttable}
\input{tables/hcn_vibron}
\begin{tablenotes}
\item[a] Computed using cc-pVTZ/0.1393-def2-QZVP*/8s8p8d8f bases and SOS$'$ scaling factors $c^\text{gs}_\text{ep}=0.91$ and $c^\text{ex}_\text{ep}=1.0$.
\item[b] Obtained from Ref. \citenum{Pavosevic_Tao_Culpitt_Zhao_Li_Hammes-Schiffer_2020},
 which used cc-pVDZ/cc-pVDZ/PB4-F2a$'$ basis sets.
\end{tablenotes}
\end{threeparttable}
\label{tab:hcn_vibron}
\end{table*}
\subsection{HNC and FHF$^-$}
\begin{figure}
    \centering
    \includegraphics[width=3.25in]{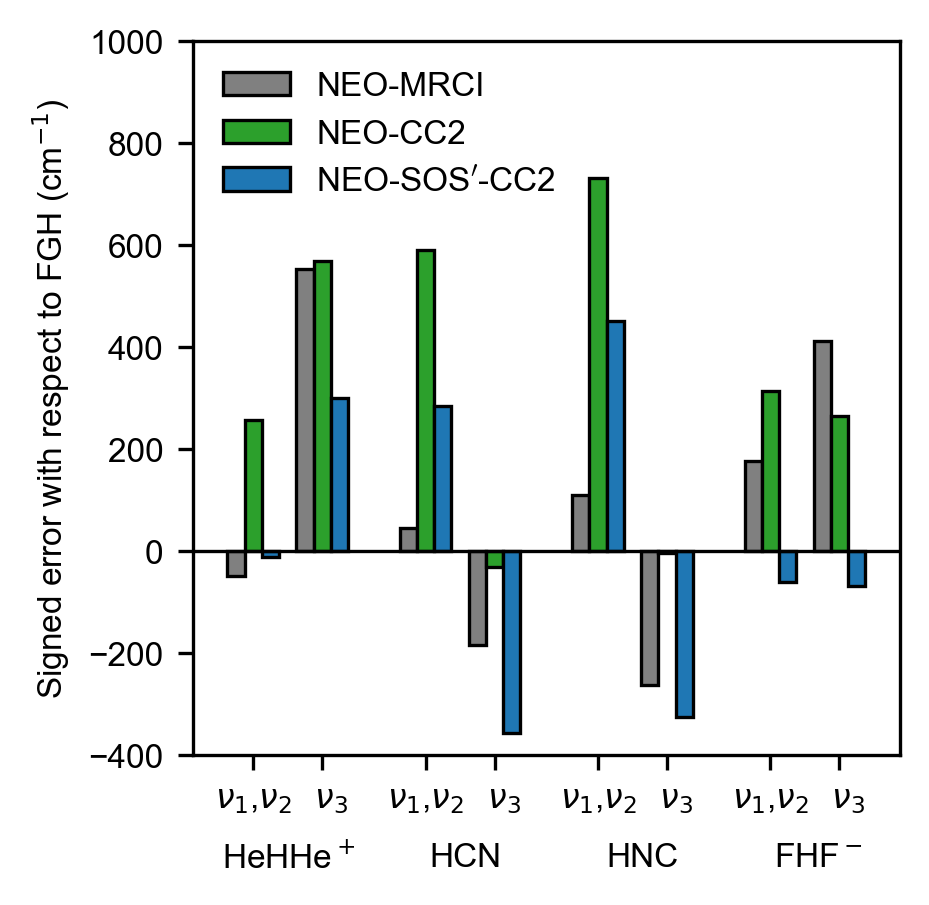}
    \caption{Signed error of the fundamental frequencies with respect to FGH values for four molecules. These frequencies correspond to the degenerate bending modes ($\nu_1$,$\nu_2$) and the stretching mode ($\nu_3$) in each molecule. The NEO-MRCI results obtained from Ref. \citenum{Malbon_Hammes-Schiffer_2025} are shown in gray, the NEO-CC2 results are shown in green, and the NEO-SOS$'$-CC2 with $c_\text{ep}^\text{gs}=0.91$ and $c_\text{ep}^\text{ex}=1.0$ results are shown in blue, appearing in this order from left to right. Note that the error of NEO-CC2 for the stretching mode of HNC is too small to be visible on this scale.}
    \label{fig:fundamentals}
\end{figure}

To obtain a better overall sense of the accuracy of NEO-SOS$'$-CC2, we investigate two additional molecules, HNC and FHF$^-$. Combined with HeHHe$^+$ and HCN, we now have a set of four molecules, two that are charged with an internal hydrogen and two that are neutral with a terminal hydrogen. This test set is the same as that used in Ref.~\citenum{Malbon_Hammes-Schiffer_2025}, allowing a direct comparison to NEO-MRCI. To facilitate this comparison, we use exactly the same basis sets and geometries. Fig. \ref{fig:fundamentals} shows the signed error with respect to the FGH reference for the fundamental frequencies of the four molecules. The errors are indicative of an incomplete basis set, as well as incomplete inclusion of electron-proton correlation. The electron-proton correlation scaling of NEO-SOS$'$-CC2 results in overall lower errors than NEO-CC2, except for the HCN and HNC stretching modes. NEO-SOS$'$-CC2 outperforms NEO-MRCI in predicting the fundamental frequencies for FHF$^-$ and HeHHe$^+$ but is less effective for HCN and HNC. Note that the electron-proton correlation scaling is likely compensating for finite basis set effects, and new scaling factors would need to be obtained for a converged basis set.
\section{Conclusion}
\label{sec:conclusion}
In this work, we introduce the NEO-CC2 and NEO-SOS$'$-CC2 methods for computing excitation energies of multicomponent systems. For the excited states of positronium hydride, NEO-CC2 is less accurate than the more expensive NEO-EOM-CCSD when benchmarked against NEO-FCI. However, NEO-SOS$'$-CC2 is able to achieve comparable accuracy to NEO-EOM-CCSD with the introduction of a scaling factor that increases the magnitude of the electron-positron correlation. This scaling factor improves both ground and excited states. For systems with quantum protons, however, scaling factors that increase the ground-state electron-proton correlation have a deleterious effect on excited states. We find that scaling factors that either decrease the ground-state electron-proton correlation or increase the excited state electron-proton correlation lead to better agreement with reference excitation energies. For all calculations on systems with quantum protons, we use fixed $c_\text{ep}$ scaling parameters optimized for a single model system with no further system-specific tuning. 

For protonic excitations, NEO-CC2 and NEO-SOS$'$-CC2 are able to capture qualitative features that have previously only been observed in the spectra of much more costly methods such as NEO-EOM-CCSD and NEO-MRCI. In particular, we observe a combination band and overtones at roughly twice the fundamental bending frequency for the molecules studied. Additionally, NEO-SOS$'$-CC2 predicts mixed electron-proton double excitations that are similar to those observed with NEO-EOM-CCSD. Both of these features are missing from all NEO-TDDFT and NEO-TDHF spectra published to date. Using the recently introduced $\gamma$-def2-QZVP* electronic basis sets, we are able to achieve qualitatively correct spectra using NEO-CC2, and in some cases approach quantitative accuracy using NEO-SOS$'$-CC2. Both NEO-CC2 and NEO-SOS$'$-CC2 will benefit from overall improvements in electronic and protonic basis sets for multicomponent excited-state methods. Moreover, NEO-SOS$'$-CC2 will benefit from more systematic tuning of the $c_\text{ep}$ parameters, particularly for more complete basis sets. Despite these challenges, the NEO-SOS$'$-CC2 approach is a promising {\it ab initio} alternative to NEO-TDDFT with the potential for identical $\mathcal{O}(N^4)$ scaling.
\section*{Acknowledgements}
The authors thank Christopher Malbon, Scott Garner, and Rowan Goudy for helpful discussions. This work was supported by the National Science Foundation Grant No. CHE-2408934. 
\section*{Supplementary Material}
The Supplementary Material is available free of charge: Programmable equations for the matrix elements of the NEO-CC2 Jacobian; protonic excitation energies using alternative electron-proton correlation scaling factors; comparison with NEO-EOM-CCSD for HCN with different conventional basis sets.
%
\section*{Data Availability Statement}
The data that support the findings of this study will be openly available upon publication at the Zenodo repository with digital object identifier \url{http://doi.org/10.5281/zenodo.17058124}.
\bibliography{cc2}
\includepdf[pages={{},1,{},2,{},3,{},4,{},5,{},6,{},7}]{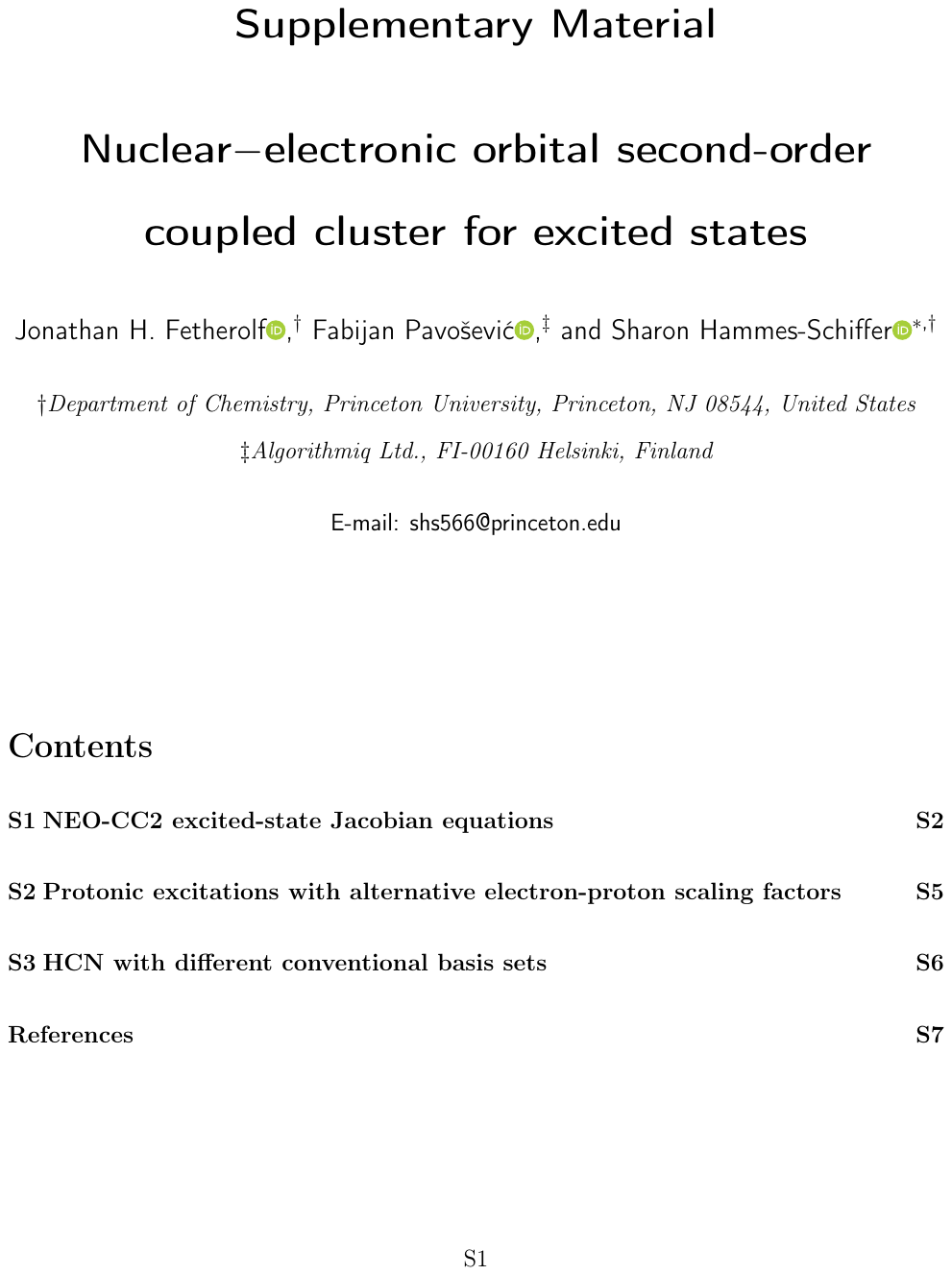}
\end{document}

%% file: tables/psh_table.tex
\begin{tabular}{ P{0.08\textwidth} l P{0.12\textwidth} P{0.1\textwidth} P{0.1\textwidth} P{0.1\textwidth} P{0.1\textwidth}}
\hline\hline
Basis & Method & G.S. $E_\text{corr}$ & 1st E.S. & 2nd E.S. & 3rd E.S. & MUE\\ \hline\hline
   \multirow{4}{*}{aTZ} & NEO-FCI$^\text{a}$ & -0.09059 & 0.1567   & 0.1659   & 0.2421 & --- \\
   & NEO-EOM-CCSD$^\text{a}$ & -0.08598 & 0.1551 & 0.1660   & 0.2396 & 0.0022$^\text{c}$ \\
   & NEO-CC2 & -0.06485 & 0.1445   & 0.1594   & 0.2415  & 0.0113$^\text{c}$ \\
   & NEO-SOS$'$-CC2$^\text{b}$ & -0.08533 & 0.1537 & 0.1684 & 0.2434 & 0.0030$^\text{c}$ \\ \hline
   \multirow{3}{*}{a5Z} & NEO-EOM-CCSD$^\text{a}$ & -0.09581  & 0.1179  & 0.1609   & 0.2244 & ---\\
   & NEO-CC2 & -0.06697  & 0.1096   & 0.1465   & 0.2198  & 0.0141$^\text{d}$ \\
   & NEO-SOS$'$-CC2$^\text{b}$ & -0.09168  & 0.1181   & 0.1573 & 0.2210 & 0.0028$^\text{d}$ \\ \hline\hline
\end{tabular}

%% file: tables/hehhe_fci.tex
\begin{tabular}{l P{0.20\textwidth} P{0.24\textwidth}}
\hline\hline
Method  & $E_0$ (Ha) & $E_1-E_0$ (cm$^{-1}$) \\ \hline
FGH$^\text{a}$  & $-$ & 771 \\
NEO-FCI$^\text{a,b}$ & -5.838083 & 783 \\ 
NEO-TDDFT$^\text{c}$ & $-$ & 588 \\ 
NEO-CC2$^\text{b}$ & -5.823354 & 1029 \\
NEO-SOS$'$-CC2 (1.3,1.3)$^\text{b}$ & -5.841000 & 2153 \\
NEO-SOS$'$-CC2 (0.91,1.0)$^\text{b}$ & -5.834333 & 767 \\
NEO-SOS$'$-CC2 (1.0,8.0)$^\text{b}$ & -5.835601 & 774 \\ \hline\hline
\end{tabular}

%% file: tables/hehhe.tex
\begin{tabular}{l P{0.15\textwidth} P{0.15\textwidth} P{0.12\textwidth} P{0.12\textwidth}}
\hline\hline
Method  & (1,0,0)/(0,1,0) & (2,0,0)/(0,2,0) & (1,1,0) & (0,0,1) \\ \hline\hline
FGH Reference$^\text{a}$  & 772  & 1584 & 1619 & 1818   \\
NEO-MR-SDT$_\text{een}$CI$^\text{a,b}$ & 723  & 1654 & 1784 & 2371   \\
NEO-TDDFT$^\text{c}$ & 611 & 5366 & 5490 & 2211 \\
NEO-CC2$^\text{b}$ & 1030 & 2407 & 2442 & 2388 \\
NEO-SOS$'$-CC2$^\text{b}$ & 762 & 2133 & 2094 & 2119 \\ \hline\hline
\end{tabular}

%% file: tables/hcn.tex
\begin{tabular}{l P{0.13\textwidth} P{0.13\textwidth} P{0.13\textwidth} P{0.13\textwidth} P{0.13\textwidth}}
\hline\hline
Method  & (1,0,0)/(0,1,0) & (1,1,0) & (2,0,0)/(0,2,0) & (0,0,1) \\ \hline\hline
FGH Reference$^\text{a}$ & 634 & 1261 & 1277 & 3173   \\
NEO-MR-SDT$_\text{een}$CI$^\text{b,c}$ & 678 & --  & --  & 2968 \\
NEO-TDDFT$^\text{d}$ & 681 & 5801 & 7262 & 3118 \\ 
NEO-CC2$^\text{c}$ & 1223 & 2970 & 3038 & 3121   \\
NEO-SOS$'$-CC2$^\text{c}$  & 917 & 2580 & 2725 & 2796 \\ \hline\hline
\end{tabular}

%% file: tables/hcn_vibron.tex
\begin{tabular}{l P{0.20\textwidth} P{0.24\textwidth}}
\hline\hline
Assignment  & NEO-SOS$'$-CC2$^\text{a}$ & NEO-EOM-CCSD$^\text{b}$ \\ \hline
$e_1p_{000}$  & 9.82 & 9.71 \\
$e_1p_{100}/e_1p_{010}$ & 14.03 & 13.92 \\ 
$e_1p_{001}$ & 14.25 & 14.12 \\
$e_1p_{110}$ & 14.38 & -- \\
$e_1p_{200}/e_1p_{020}$ & 14.39 & -- \\ \hline\hline
\end{tabular}